\documentclass[a4paper,11pt]{article}
\usepackage{amsfonts,amsmath,amssymb,bm,epsfig,url}
\usepackage[dvips]{color}

\usepackage{dcolumn}
\usepackage[pagebackref=false, colorlinks=true]{hyperref}
\usepackage{textcomp}
\usepackage{float}
\usepackage{subfigure}
\usepackage{graphicx}

\definecolor{redish}{rgb}{0.7,0.2,0.0}  
\definecolor{bluish}{rgb}{0.2,0.5,0.8}

\hypersetup{linkcolor=redish,          
                  citecolor=blue,        
                  filecolor=magenta,      
                  urlcolor=bluish}          

\DeclareFontFamily{U}{rsfs}{}         
\DeclareFontShape{U}{rsfs}{m}{n}{<5> rsfs5 <6><7> rsfs7          %
  <8><9><10><10.95><12><14.4><17.28><20.74><24.88> rsfs10}{}     %
\DeclareMathAlphabet{\mathfs}{U}{rsfs}{m}{n} 

\newcommand{\df}[2]{ \frac{\partial {#1}}{\partial {#2}} }

\def \O{\Omega}

\def \th{\theta}

\def \f{\frac}

\title{Gravitomagnetic effect in magnetized neutron stars}
\author{Debarati Chatterjee\footnote{dchatterjee@lpccaen.in2p3.fr}  
\\ {\it LPC/ENSICAEN, 6 Boulevard Mar\'echal Juin, 14050 Caen, France}
\\
{}
\\
Chandrachur Chakraborty\footnote{chandrachur.chakraborty@tifr.res.in (Corresponding author)}
\\{\it Tata Institute of Fundamental Research, Mumbai 400005, India}
\\
{}
\\
Debades Bandyopadhyay\footnote{debades.bandyopadhyay@saha.ac.in}
\\{\it Astroparticle Physics and Cosmology Division,}
\\{\it Saha Institute of Nuclear Physics, HBNI, Kolkata 700064, India}
}
\date{}

\begin{document}

\maketitle

\abstract{Rotating bodies in General Relativity produce frame dragging,
also known as the {\it gravitomagnetic effect} in analogy with classical 
electromagnetism. 
In this work, we study the effect of magnetic field on 
the gravitomagnetic effect in neutron stars 
with poloidal geometry, which is produced as a result of its rotation. 
We show that the magnetic field has a non-negligible impact on frame dragging. 
The maximum effect of the magnetic field appears along the polar direction,
where the frame-dragging frequency decreases with increase in magnetic field, 
and along the equatorial direction, where its magnitude increases.
For intermediate angles, the effect of the magnetic field decreases, 
and goes through a minimum for a particular angular value at which 
magnetic field has no effect on gravitomagnetism. Beyond
that particular angle gravitomagnetic effect increases with
increasing magnetic field. We try to identify this `null region'
for the case of magnetized neutron stars,
both inside and outside, as a function of the magnetic field,
and suggest a thought experiment to find the null region
of a particular pulsar using the frame dragging effect.
}

\section{Introduction}
\label{secintro}
Neutron stars are ideal testing grounds for studying 
strong gravitational effects as predicted by General Relativity. 
Frame-dragging \cite{hp} is one such important relativistic effect 
in compact stars. General relativity predicts that the curvature of spacetime
is produced not only by the distribution of mass-energy
but also by its motion. Similarly, in electromagnetism, the fields
are produced not only by charge distributions but also 
by currents, i.e., as the moving charge produces the magnetic field.
Considering this analogy, this effect is also called the 
{\it gravitomagnetic effect}. The rotation of a massive body 
distorts the spacetime \cite{lt}, making the spin of nearby test 
gyroscope precess \cite{schiff}. It has been pointed out in Ref. \cite{hp}
that most of the merit for the discovery of frame-dragging belongs to Einstein
but in the existing literature this is referred as Lense-Thirring (LT) effect.

Frame-dragging is generally studied in the context of the accretion disc theory where
the orbital plane precession frequency is calculated for a {\it test particle} 
orbiting in the equatorial plane relative to the asymptotic observer.
In this article, we derive the frame-dragging precession frequency or
LT precession frequency of a test gyroscope ({\it spinning test particle})
relative to the ``fixed stars'' (Copernican system),
which was first derived by Schiff \cite{schiff} for the weak field regime.
This effect has also been derived using the Kerr metric
\cite{cm} without invoking the weak field
approximation and it has been shown that the exact formulation reduces to the 
well-known LT frequency in weak gravity regime, derived in \cite{schiff}.
This has been experimentally confirmed by the Gravity Probe B \cite{ev} in the field of 
the earth and the precession of its orbit has also been measured 
precisely by the LAGEOS satellite \cite{nat}.

The prospect of exploring frame-dragging in the field of a rotating
neutron star is attractive, as LT precession frequencies
there can be larger than those caused by classical oblateness effects
\cite{Morsink}. Further, 
as the frame dragging effect is also found to occur inside rotating neutron stars and 
the magnitude of this effect depends on the density profile, it is an 
interesting prospect to try to constrain the equation of state exploiting 
this effect \cite{Morsink,Kalogera}.
The matter orbits in accretion disks can provide ways to probe the spacetime
properties in the vicinity of neutron stars, including frame dragging effects. 
Quasi-periodic oscillations (QPOs)
observed in X-ray Bursts and diagnostics of X-ray spectra have been suggested as promising tools
to investigate the motion of matter around neutron stars \cite{Stella}.
Recently, it has been predicted \cite{rp} that frame-dragging can be generated 
by the electromagnetic fields. 

Using the slow-rotation approximation proposed in \cite{hartle}, the 
frame dragging effect inside a slowly rotating neutron star can 
be estimated. In this approximation, to second order in rotation 
frequencies, the structure of a star changes only by quadrupole terms 
and the equilibrium equations describing the structure reduce to a 
set of ordinary differential equations. The frame-dragging frequency
in this scheme is only a function of the radial distance from the 
centre of the star. The precession frequency at the centre is then given
by the frequency of rotation of the neutron star, and
drops off monotonically towards the surface. 
Recently some of the authors of this article derived the exact LT frequency 
of a test gyro \cite{chandra}. 
The frame dragging rate
was found to be not only a function of the rotation frequency 
of the star but also of the other metric components. Unlike in the 
slow rotation limit, it was found that
the precession rate depends both on the radial distance ($r$) 
and the colatitude ($\theta$) of the gyroscope. Thus, in rapidly rotating
neutron stars, the precession rates are not found to be the
same in the equatorial and polar directions.

It is well known that neutron stars are endowed with strong magnetic fields. 
The surface magnetic field value for normal pulsars, estimated from their spin down rates,
points to values around $10^{12} - 10^{13}$ G. There have been reports of discovery of high magnetic
field neutron stars, both in isolated systems (XDINSs, RRATs) \cite{Olausen, Kaspi} 
as well as in binaries (SXTs) \cite{Ho}. The largest magnetic fields have been observed in 
Anomalous X-ray Pulsars (AXPs) and Soft Gamma-ray Repeaters (SGRs), commonly called magnetars.
Various observations, including direct detection of cyclotron features
in the spectra \cite{Ibrahim, Mereghetti}
confirm that these objects possess surface magnetic fields as large as
$10^{15} - 10^{16}$ G. However, magnetic
fields in the interior of magnetars could be even larger. As it cannot
be directly measured, the maximum interior
magnetic field is usually estimated using the scalar virial theorem, which
points towards a value of $10^{18}$ G.
The presence of a strong magnetic field breaks of spherical
symmetry of the star, resulting in a strong deformation
of its shape from isotropy. The investigation of the effect of magnetic field on the 
frame-dragging rate in neutron stars is the aim of this paper.

The scheme of the paper is as follows.
In Sec. \ref{seceq} of this paper, we define the basic equations to describe 
the exact frame dragging rate in rotating neutron stars 
in presence of an electromagnetic field. Sec. \ref{secnum} explains 
the numerical scheme for solving the equations described in the previous
section. The results obtained are discussed in Sec. \ref{secres} and Sec. \ref{secnull}
is devoted to suggest a thought experiment to find 
the `null region' of a pulsar using LT precession.
Finally Sec. \ref{seccon} closes the article with a summary and future outlook.

\section{Formalism}
\label{seceq}
\subsection{Global models of rotating neutron stars}
\label{sec:global}

To construct numerical models of rotating neutron stars endowed
with a magnetic field within a framework of general relativity,
we follow the approach of \cite{BGSM}. We recapitulate here the main assumptions and steps
to construct global models. 

 To construct the simplest model of a magnetized neutron star, we restrict ourselves to stationary 
and axisymmetric configurations, where the magnetic axis is aligned with the rotation axis.
We also assume that the star does not undergo meridional circulation, i.e.
the spacetime is circular.

 With the assumption of a stationary, axisymmetric spacetime, in which the matter part fulfills 
the circularity condition, the line element can be written as \cite{chatterjee}:
\begin{equation}
\label{eq:metric}
ds^2 = - N^2 dt^2 + A^2(dr^2+r^2 d\theta^2) + B^2 r^2 \sin^2 (d\phi - N^{\phi} dt)^2
\end{equation}
where $N,N^{\phi},A,B$ are functions of $(r,\theta)$. 

In the same way as in~\cite{Bocquet} we consider here that the
electromagnetic field originates from an electric current distribution, denoted hereafter
simply by $j^\sigma$, which are \textit{a priori} independent from the
movements of inert mass (with 4-velocity $u^\mu$). 
This assumption is justified because a neutron star is made of neutrons which do not carry any charge and, therefore, 
charged currents are produced by other particles, which carry less mass. Ideally, 
one should write down a description using multi-fluid MHD (Magnetohydrodynamics) (see e.g. \cite{Spruit2016}), with
one fluid carrying no charge (neutrons), and another fluid carrying charged currents. 
However using this simplified model, we are able to generate an electromagnetic field simply by adjusting the free current.

If the magnetic field possesses a mixed poloidal-toroidal geometry, or meridional circulation of matter, the circularity property
of the spacetime would be broken and would render the construction of global models much more complicated \cite{EricBonazzolla93}.
Therefore for a circular spacetime, we can have only purely poloidal or 
toroidal magnetic field configuration.

For this study, we consider only a purely poloidal magnetic field configuration.
The choice of poloidal magnetic field geometry is motivated by the fact that it is the surface poloidal
magnetic field which can be estimated from the observed spin-down of
pulsars via electromagnetic radiation, assuming a simple dipole model. 
The virial theorem allows to estimate a
maximum limit to the magnetic field in the interior, irrespective of
the configuration, from the observed poloidal surface fields. Although the
study of magnetized neutron star models with purely poloidal magnetic
field is not the most general one, it gives us a fairly good idea
about the effect of the maximum field on the stellar structure.
We discuss later in Sec. \ref{sec:geometry} the limitations of 
such an assumption.

It was demonstrated by Carter \cite{Bocquet} that the most general form of the electric four-current
compatible with the assumptions of stationarity, axisymmetry and circularity has the form $j^{\alpha} = (j^t,0,0,j^{\phi})$.
Then the electromagnetic field must be deduced from a potential $A_{\alpha} = (A_t,0,0,A_{\phi})$ using $F_{\alpha \beta} =
A_{\beta, \alpha} - A_{\alpha, \beta}$. 
 The electric and magnetic
fields measured by the Eulerian observer (whose four-velocity is
$n^\mu$) are then defined as $E_\mu = F_{\mu\nu}\, n^\nu$ and $B_\mu =
-\frac{1}{2} \epsilon_{\mu\nu\alpha\beta}\, n^\nu\, F^{\alpha\beta}$,
with $\epsilon_{\mu\nu\alpha\beta}$ is the Levi-Civita tensor associated
with the metric~(\ref{eq:metric}). The non-zero components read
(see \cite{chatterjee,Bocquet} for detailed calculations):
  \begin{eqnarray}
    E_r & = & \frac{1}{N} \left( \df{A_t}{r} + N^\varphi
      \df{A_\varphi}{r} \right)\\
    E_\theta & = & \frac{1}{N} \left( \df{A_t}{\theta} + N^\varphi
      \df{A_\varphi}{\theta} \right) \\ 
    B_r & = &
    \frac{1}{Br^2\sin\theta}\df{A_\varphi}{\theta} \\
    B_\theta & = & - \frac{1}{B \sin \theta} \df{A_\varphi}{r} .
  \end{eqnarray}
The Maxwell equations can then be obtained from the relation 
$F^{\alpha \beta}_{;\beta} = \mu_0 j^{\alpha}$,
 where $\mu_0$ is the permeability of free space.

The energy-momentum tensor in the presence of a magnetic field can be written as:
\begin{eqnarray}
T^{\mu \nu} &=& (\varepsilon + P) u^{\mu} u^{\nu} + P g^{\mu \nu}
- \frac{1}{\mu_0} \left(F^{\mu \alpha} F^{\nu}_{\alpha} + \frac{g^{\mu \nu}}{4} F_{\alpha \beta} 
F^{\alpha \beta}\right) ~.
\label{emtensor}
\end{eqnarray}
Here the first term represents the matter contribution (perfect fluid), where $\varepsilon$ is the energy density, 
$P$ is the pressure and $u^{\mu}=(u^t,0,0,u^{\phi})$ is the fluid four velocity. The last term 
gives the electromagnetic field contribution to the energy-momentum tensor.
It was shown in \cite{chatterjee} that the contributions of magnetization and the magnetic field dependence of the Equation of State (EoS)
on the structure of magnetized neutron stars are negligible, hence we disregard 
them in the calculations. 

In presence of an electromagnetic field, the equilibrium configurations for 
rotating neutron stars are determined by solving the coupled
Maxwell-Einstein equations $$\nabla_{\mu} T^{\mu \nu}=0.$$ In the case of rigid rotation, this is equivalent 
to solving a first integral of the equation of fluid stationary motion 
\cite{BGSM,chatterjee,Bocquet}:
\begin{equation}
\label{eq:first-integral} 
H(r,\theta) + \nu(r,\theta) - \ln \Gamma (r,\theta) + \Phi(r,\theta) = \text{const} ~.
\end{equation}
In the above equation:\\
$H = {\rm ln} \left(\frac{e+p}{n m_B c^2}\right)$ is the fluid log-enthalpy, where 
$e$ is the proper energy density of the fluid, $p$ is the pressure, $n$
the baryon density, $m_B$ the mean baryon rest mass;
$ \nu = {\rm ln} N$ is the gravitational potential related to the metric component
$N$ and $ \Gamma$ is the Lorentz factor relating the Eulerian observer to the fluid comoving observer.\\
We take into account the Lorentz force exerted by the 
electromagnetic field on the medium via the electromagnetic term:
\begin{equation}
\label{eq:em-term}
\Phi(r,\theta) = - \int_0^{A_{\phi}(r,\theta)} f(x) dx
\end{equation}
where $f$ is an arbitrary current function relating the components of the 
electric current to the electromagnetic potential $A_{\phi}$:
\begin{equation}
j^{\phi} - \Omega j^t = (e+p)f(A_{\phi})~.
\end{equation}
The constant current function is a simple
toy model to generate efficiently a dipolar magnetic field with the maximum at the centre
of the neutron star.

\subsection{Calculation of the frame dragging rate}
\label{sec:ltprec}

The exact frame dragging rate in a rotating neutron star was calculated for a stationary, axisymmetric
and circular metric in  \cite{chandra}. Here we do not recapitulate the derivation and urge the 
interested reader to refer to \cite{chandra} for the details. In this work, we simply rewrite
the precession frequencies in terms of the metric Eq. (\ref{eq:metric}) used to construct the global models
for strongly magnetized neutron stars (Sec. \ref{sec:global}).

In the orthonormal 
coordinate basis, the exact LT precession rate of 
a test gyro relative to the Copernican system (or ``fixed stars'')
for the rotating neutron star with electromagnetic field can be written as:
\begin{eqnarray}\nonumber
\label{eq:omega_lt}
&& \vec{\O}_{LT}=\f{1}{2NA(-N^2+(N^{\phi})^2r^2B^2\sin^2\th)}. 
\\
&& \left[\sin\th[N^2(BrN^{\phi}_{,r}+2N^{\phi} rB_{,r}+2N^{\phi} B)
+(N^{\phi})^2 r^3B^3N^{\phi}_{,r}\sin^2\th-2N^{\phi} BrNN_{,r}]\hat{\th}\right.
\label{ltnp} \nonumber
\\ 
&&
\left.-[N^2(BN^{\phi}_{,\th}\sin\th+2N^{\phi} B_{,\th}\sin\th +2N^{\phi} B\cos\th)
+(N^{\phi})^2 r^2B^3N^{\phi}_{,\th}\sin^3\th-2N^{\phi} BNN_{,\th}\sin\th]\hat{r}\right]~.\nonumber
\\
\end{eqnarray}

The modulus of Eq. (\ref{eq:omega_lt}) is
\begin{eqnarray}\nonumber
\label{eq:omega_lt_mod}
&&\O_{LT}=|\vec{\O}_{LT}(r,\th)| 
\\&=&\f{1}{2NA(-N^2+(N^{\phi})^2r^2B^2\sin^2\th)}. \nonumber
\\
&&\left[\sin^2\th[N^2(BrN^{\phi}_{,r}+2N^{\phi} rB_{,r}+2N^{\phi} B)
+(N^{\phi})^2 r^3B^3N^{\phi}_{,r}\sin^2\th-2N^{\phi} BrNN_{,r}]^2\right. \nonumber
\\ 
&&
\left.+[N^2(BN^{\phi}_{,\th}\sin\th+2N^{\phi} B_{,\th}\sin\th +2N^{\phi} B\cos\th)
+(N^{\phi})^2 r^2B^3N^{\phi}_{,\th}\sin^3\th-2N^{\phi} BNN_{,\th}\sin\th]^2\right]^\f{1}{2}
\nonumber
\\
\end{eqnarray}
where `$,r$' and `$,\theta$' represent the derivatives with respect to $r$ and $\theta$ respectively.

We must note here that the above expression depends only on the metric coefficients, and not explicitly on
the magnitude or geometry of the magnetic field.

\section{Numerical scheme}
\label{secnum}

In general the Einstein-Maxwell equations described above are fourth-order coupled differential equations 
which are challenging to solve numerically. However if the four-dimensional spacetime is foliated using the 3+1 formalism (where
spacetime is split into three dimensional hypersurfaces evolving in time), they reduce to Poisson-like partial differential equations.
These equations can then be solved iteratively, each step consists of solving Poisson equations using a Chebyshev-Legendre 
spectral scheme. This is essentially what has been done within the numerical library LORENE \cite{grandclement-09}.
The input parameters for the model are an EoS, the current function $f$, the rotation frequency $\Omega$ and the central log-enthalpy $H_c$. The electromagnetic
field originates from the free currents, which are controlled via the current function $f$. Varying the input current function $f$ allows us to control the 
intensity of the magnetic field $B$ or the magnetic moment $\cal{M}$. On the other hand, varying the central log-enthalpy permits variation of the mass of the neutron star 
(gravitational mass $M_G$ or baryon mass $M_B$).  

\begin{figure}
\begin{center}
\includegraphics[width=.5\textwidth,angle=270]{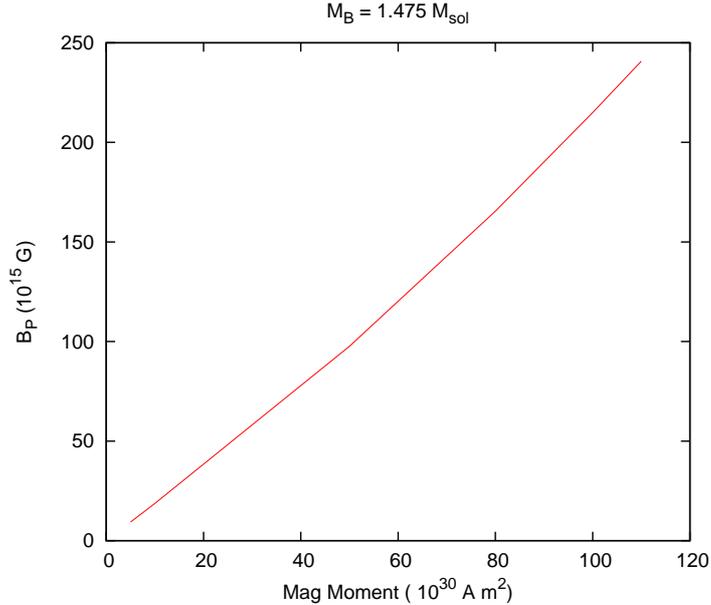}
\caption{Magnetic moment vs polar magnetic field ($B_P$) for a non-rotating pulsar having a baryonic mass $M_B$ of 1.475 M$_{\odot}$.}
    \label{fig:bpmm}
    \end{center}
  \end{figure}
  
Using the numerical 
scheme described above, we compute models of fully relativistic neutron stars for
a given magnetic field configuration. We consider a poloidal magnetic field configuration.
We choose as an example a particular pulsar PSR J0737-3039
whose mass ($1.337 M_{\odot}$) and spin ($\Omega=276.8$ s$^{-1}$) are well known.
The exact frame-dragging rate for this pulsar was calculated for the non-magnetic case in \cite{chandra}.
We now study the effect of magnetic field on its frame dragging rate.
Just as in the case of a rotating star it is not the rotation frequency but the moment of inertia which is a constant of motion,
for magnetized neutron stars the magnetic field is not constant of motion but rather the magnetic moment. 
To give a clearer correspondence between the polar magnetic field of the pulsar and its magnetic moment, we plot in  Fig. \ref{fig:bpmm}
the relation between the two for a given baryonic mass of 1.475 $M_{\odot}$, which corresponds to a gravitational mass
of  1.337 $M_{\odot}$ for the zero magnetic field case. We can see from the figure that the relation is almost linear,
with the largest value $1.1 \times 10^{32}~ A.m^2$ for the magnetic moment corresponding to a magnetic field value of 
$2.4 \times 10^{17}$ G.

\begin{figure}[tbp]
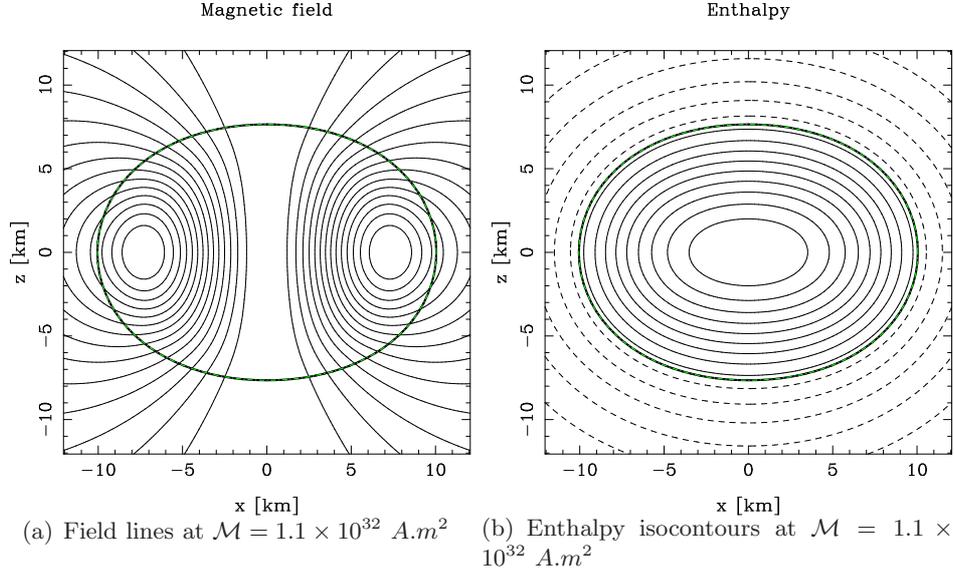

\centering
\subfigure[Field lines at ${\cal{M}}=1.1 \times 10^{32}~ A.m^2$]{
     \includegraphics[width=.4\textwidth,angle=270]{field_mm1.1e32.eps}}
\subfigure[Enthalpy isocontours at ${\cal{M}}=1.1 \times 10^{32}~ A.m^2$]{
      \includegraphics[width=.4\textwidth,angle=270]{enthalpy_mm1.1e32.eps}}
\caption{ Magnetic field lines (Panel (a)) and enthalpy
isocontours (Panel (b)) in the meridional plane
for PSR J0737-3039, for a magnetic moment of $1.1 \times 10^{32}~ A.m^2$}
\label{fig:xzcut}
  \end{figure}

To show the effect of magnetic field on the structure of a neutron star,
we demonstrate an equilibrium neutron star configuration
corresponding to a large magnetic moment of $1.1 \times 10^{32}$
A m$^2$. In Fig. \ref{fig:xzcut}, the magnetic field lines (Panel (a)) and 
enthalpy isocontours (Panel (b)) in the meridional plane
for PSR J0737-3039, having a gravitational mass of 1.337 M$_{\odot}$ and angular frequency of 276.8 s$^{-1}$,
for a magnetic moment of $1.1 \times 10^{32}~ A.m^2$ have been displayed. It is evident from the figure, that under the influence
of the strong magnetic field, the stellar surface has been highly deformed 
from spherical symmetry. 

To estimate the deformation of the structure due to the magnetic field, we compared the values of 
the $r_{pole}/r_{eq}$ and the quadrupole moment of the star due to different constant magnetic
moments. The values are displayed in Table \ref{tab:deform}. It is evident from the table that with increase
in magnetic moment, the flattening increases i.e., the ratio of polar radius with respect to the equatorial radius decreases
from 1 (for a spherical star). The equatorial circumferential radius as well 
as the quadrupole moment increase with increase in magnetic field.

\begin{table*}
 \begin{minipage}{140mm}
  \caption{Deformation of magnetized neutron stars for varying magnetic moments $\cal{M}$ (in A.m$^2$). 
  Displayed below are the 
  values of polar to equatorial flattening $r_{pole}/r_{eq}$, corresponding equatorial circumferential radii
  $R_{circ}$ (in km) and the quadrupole moments $Q$ (in $10^{38}$ kg.m$^2$).}
   \begin{center}
 \begin{tabular}{@{}|l|r|r|r|@{}}
\hline
  {$\cal{M}$} & $r_{pole}/r_{eq}$ & $R_{circ}$ & $Q$\\
      (A.m$^2$)  & {} & (km) & ($10^{38}$ kg.m$^2$) \\
 \hline
 $5\times10^{30}$ & 0.998 & 11.355 & 0.00067 \\
        $10^{31}$ & 0.997 & 11.361 & 0.00197 \\
 $5\times10^{31}$ & 0.937 & 11.540 & 0.04088 \\ 
 $8\times10^{31}$ & 0.856 & 11.815 & 0.09581 \\
        $10^{32}$ & 0.793 & 12.054 & 0.14013 \\
$1.1\times10^{32}$ & 0.761 & 12.189 & 0.16399 \\
\hline
\end{tabular}
\end{center}
\label{tab:deform}
\end{minipage}
\end{table*}

\section{Results and Discussions}
\label{secres}
An EoS of neutron star matter is needed for this calculation. The recently 
observed 2M$_{\odot}$ neutron star puts a strong constraint on the EoS 
\cite{watts}. In an earlier calculation of the LT precession rate without 
magnetic field \cite{chandra}, we adopted three different EoSs based on the
model of Akmal, Pandharipande and Ravenhall (APR), chiral model and density
dependent (DD) relativistic mean field model. All three EoSs satisfy the two 
solar mass neutron star constraint. It was shown in \cite{chandra} that the 
behaviour of the LT-precession rate for different EoSs is qualitatively similar. 
Therefore in this study, we employ only the APR \cite{APR} EoS. The maximum 
masses as well as the corresponding radii of static neutron stars as well as
those rotating at Kepler frequency were displayed already in
\cite{chandra}, hence we do not repeat this study here.
 Our aim is to perform a study of the effect of magnetic field
on the frame dragging rate $\Omega_{LT}$ for neutron stars
rotating at sub-Kepler frequencies.

We construct equilibrium configurations of neutron stars for
constant values of magnetic moment $\cal{M}$ = $5 \times 10^{30}, 10^{31}, 5 \times 10^{31}, 
8 \times 10^{31}, 10^{32}, 1.1 \times 10^{32}~ A.m^2$.
For a given magnetic moment, we calculate the value of the frame-dragging precession rate 
inside ($r \leq r_s$) as well as outside the pulsar 
($r>r_s$), where $r_s$ is the distance of the surface of the pulsar
from its centre (at the equator $r_s=r_{eq}$; $r_{eq}$ is the equatorial coordinate radius). 

We should clarify that the LT frequency can be calculated using Eq. (\ref{eq:omega_lt_mod})
both inside and outside of a pulsar. As all the parameters ($A, B, N, N^{\phi}$)
in Eq. (\ref{eq:omega_lt_mod}) are functions of $r$ and $\theta$, the equation 
is automatically modified by changing the values of $r$ and $\theta$. 
Inside the pulsar ($r \leq r_s$) all of the above mentioned parameters
take the values corresponding to the solution of the Einstein equation
with the energy-momentum tensor mentioned in Eq. (\ref{emtensor}).
Outside the pulsar, ($r > r_s$) Eq. (\ref{emtensor}) reduces to the case with 
$\varepsilon=P=0$ in absence of matter and Eq.(\ref{eq:metric})
describes the electrovacuum solution of the Einstein equation.
Thus, the frame-dragging formula (Eq. (\ref{eq:omega_lt_mod}))
 also gets modified automatically with the corresponding values of $A, B, N, N^{\phi}$
and the LT precession rates (for a given angle) are continuous from the centre to
the asymptotically large distances as well as on the surface of the pulsar.

One must note that the effect of magnetic field on the frame-dragging frequency is two fold:\\
(i) Firstly, there is the effect of electromagnetic field directly 
on the matter and vacuum solutions, given by Eq. (\ref{eq:omega_lt_mod}). \\
(ii) Secondly, magnetic field causes a deformation of the stellar surface.
Thus it affects the radial distance
at which the energy density falls to zero, and consequently the LT 
frequency also shifts at the radial distance at
which the matter solution matches the electrovacuum
solution.

\subsection{Interior region ($r \leq r_s$)}
\begin{figure}[tbp] 
\begin{center}
\subfigure[at ${\cal{M}}=5 \times 10^{30}~ A.m^2$]{
      \includegraphics[width=.4\textwidth,angle=270]{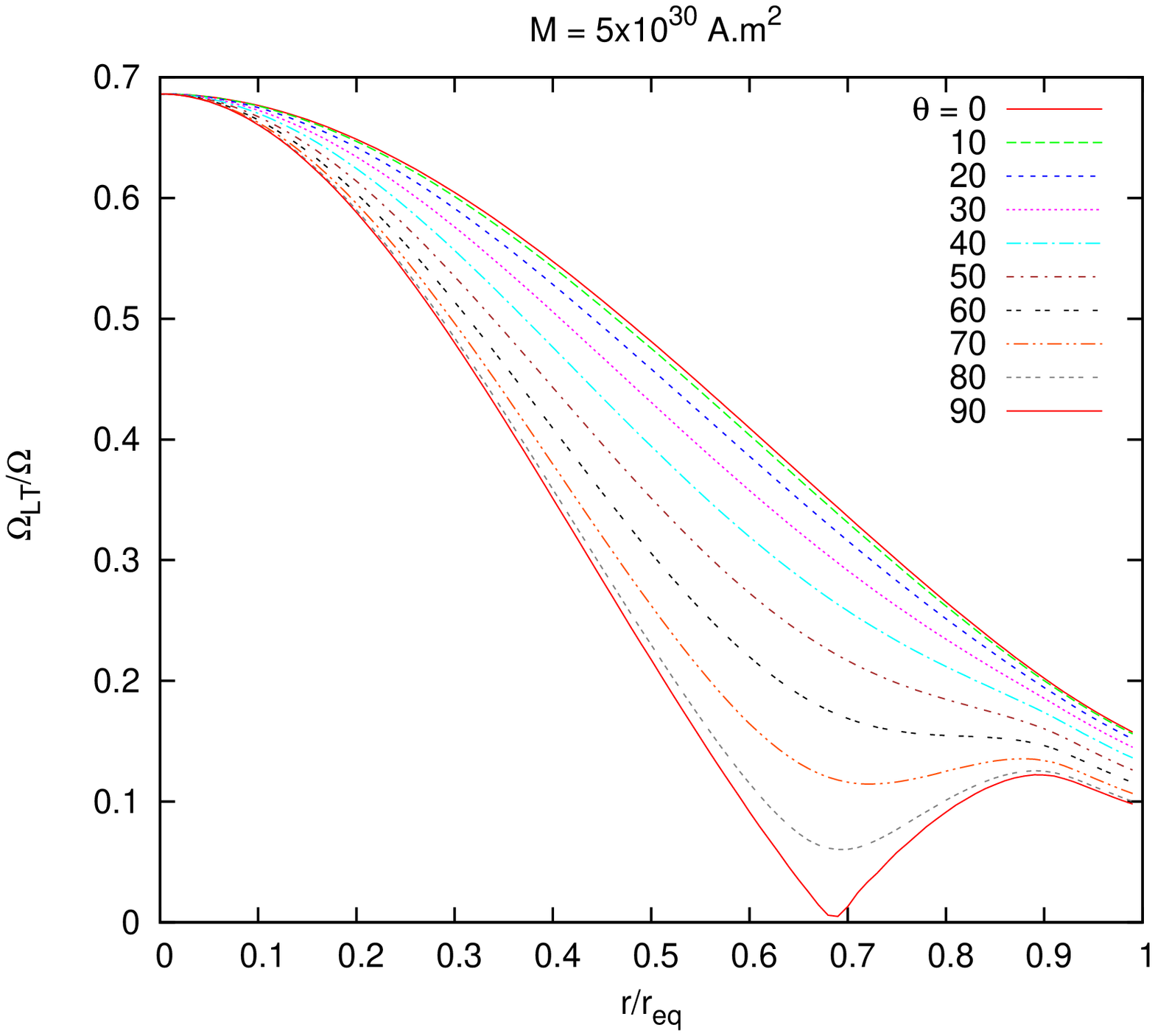}}
\subfigure[${\cal{M}}=1.1 \times 10^{32}~ A.m^2$]{
     \includegraphics[width=.4\textwidth,angle=270]{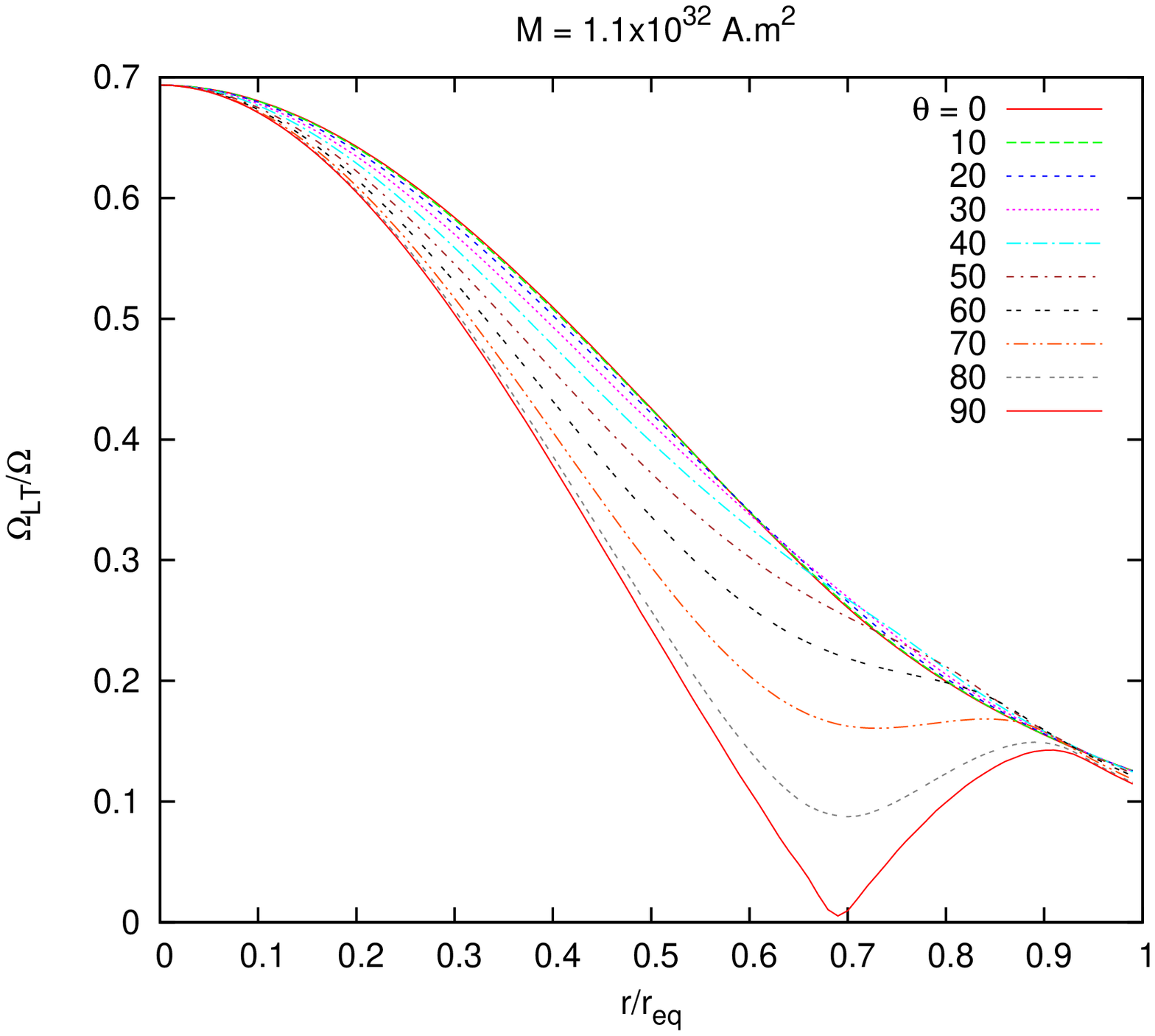}}
\caption{Normalized frame-dragging rate $\Omega_{LT}/\Omega$ inside the pulsar as a function
of radial distance $r$ in units of $r_{eq}$ along the pole 
($\theta = 0^0$) to equator ($\theta = 90^0$ ) 
for two values of magnetic moment $5 \times 10^{30}$ and $1.1 \times 10^{32}~ A.m^2$.}
    \label{fig:in}
    \end{center}
  \end{figure}

In Panel (a) and Panel (b) of Fig. \ref{fig:in},
we display the frame dragging rate as a function of radial distance ($r/r_{eq}$)
for different angular values from 0 (along polar direction) to 90 (equatorial direction)
for two different constant magnetic moments. 
The results are qualitatively the same as obtained in 
\cite{chandra}: the frame dragging frequency has a smooth
behaviour along the pole from the centre 
to the surface of the star but the behaviour is quite 
different along the equator, which is already been explained
in \cite{chandra} and \cite{cc2}.
Along the equatorial direction the frame dragging frequency goes through a ``dip'' or local minimum
at a certain angle, labeled as ``critical angle".
In this case, the critical angle occurs around $\sim 60^0$.
For magnetic moments ${\cal{M}} \le 10^{31}$ A m$^2$, 
the results are indistinguishable from the non-magnetic case. It can also be 
seen from Panel (a) and Panel (b)
of Fig.\ref{fig:in} that frame-dragging rate is quite large $(\sim70\%$
of the total spin of the pulsar)
at the centre of the pulsar, where the mass-energy density is largest,
and is decreasing in the radial outward direction. This result is compatible
with the earlier result obtained in \cite{chandra}. Actually, the frame-dragging 
rates inside of the pulsars can vary in a wide range
depending on their masses, spins and mass-energy densities \cite{weber}.

\begin{figure}[tbp]
 \begin{center}
\subfigure[at $0^0$ (along the pole)]{
\includegraphics[width=2.3in,angle=270]{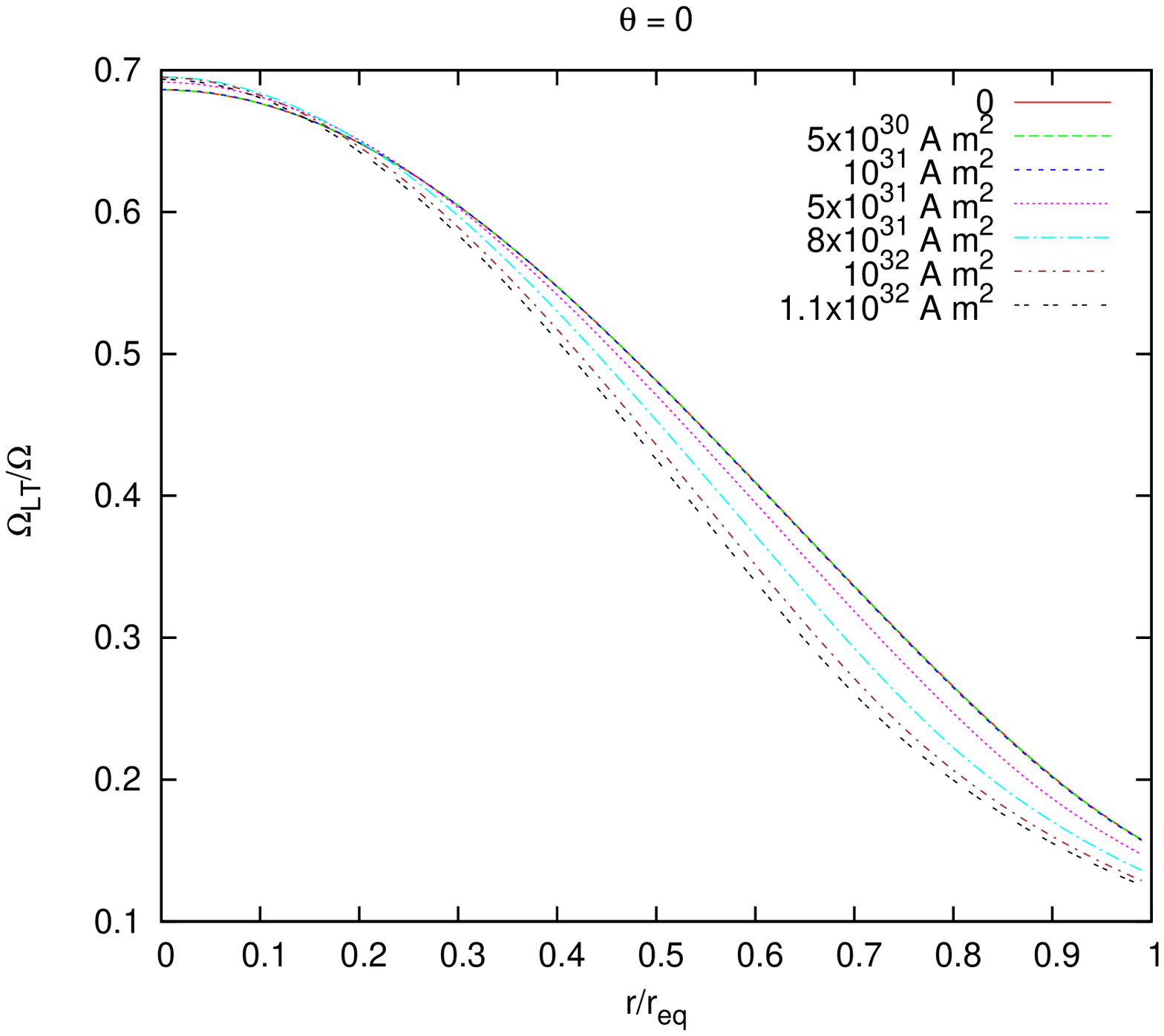}}
 \subfigure[at $20^0$ ]{
 \includegraphics[width=2.3in,angle=270]{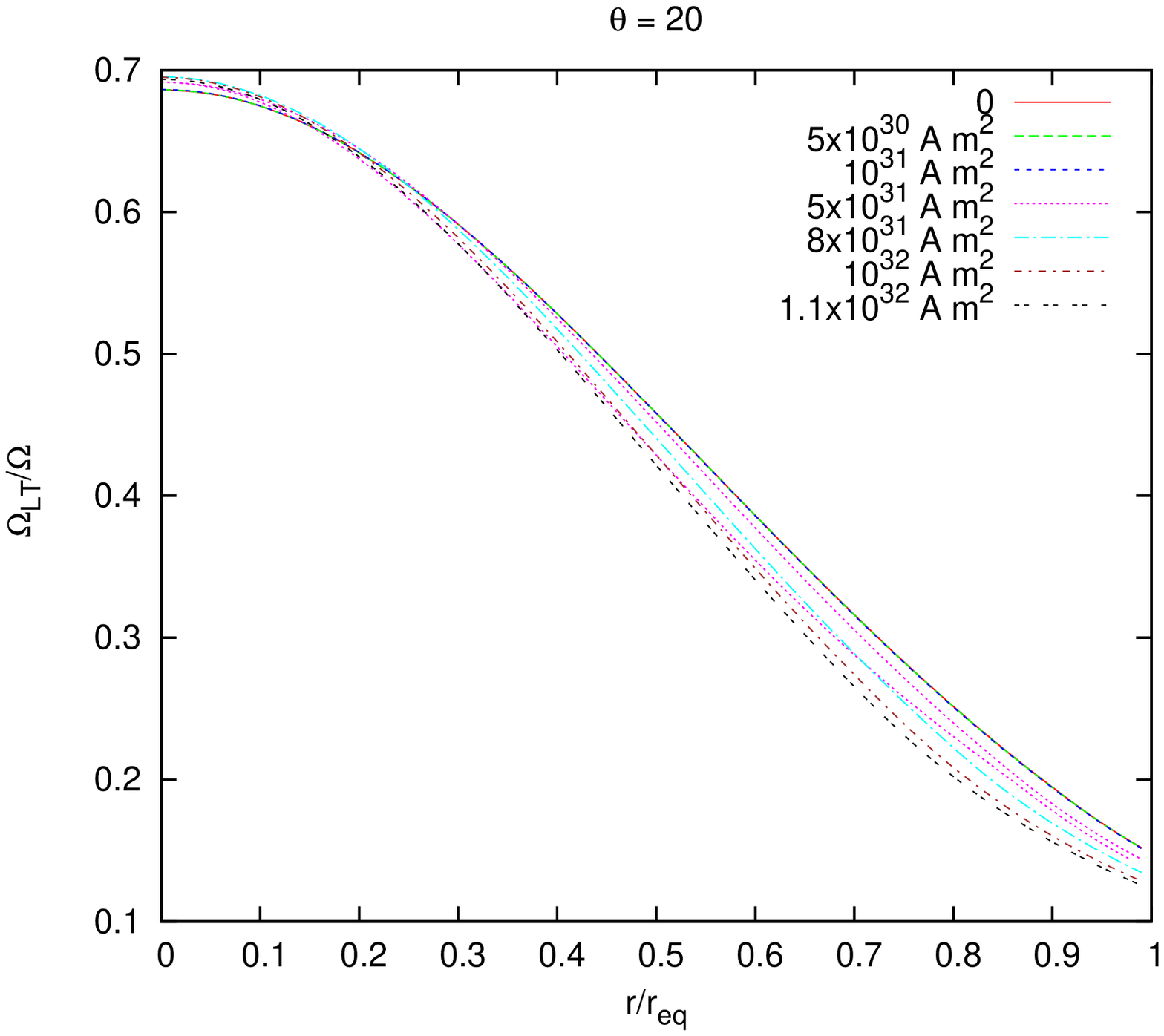}}
  \subfigure[at $40^0$ (along the null Region) ]{
 \includegraphics[width=2.3in,angle=270]{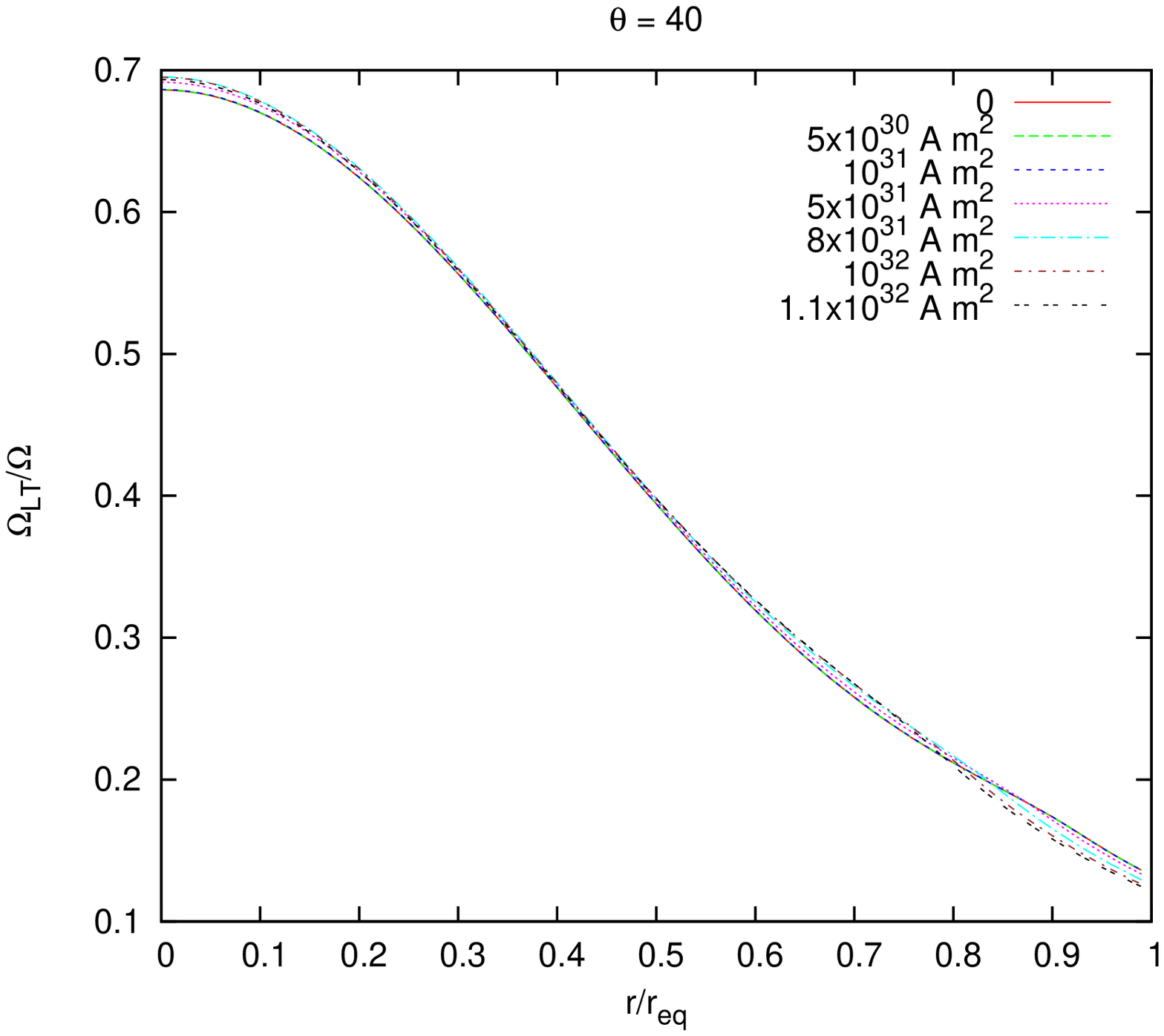}}
 \subfigure[at $50^0$]{
 \includegraphics[width=2.3in,angle=270]{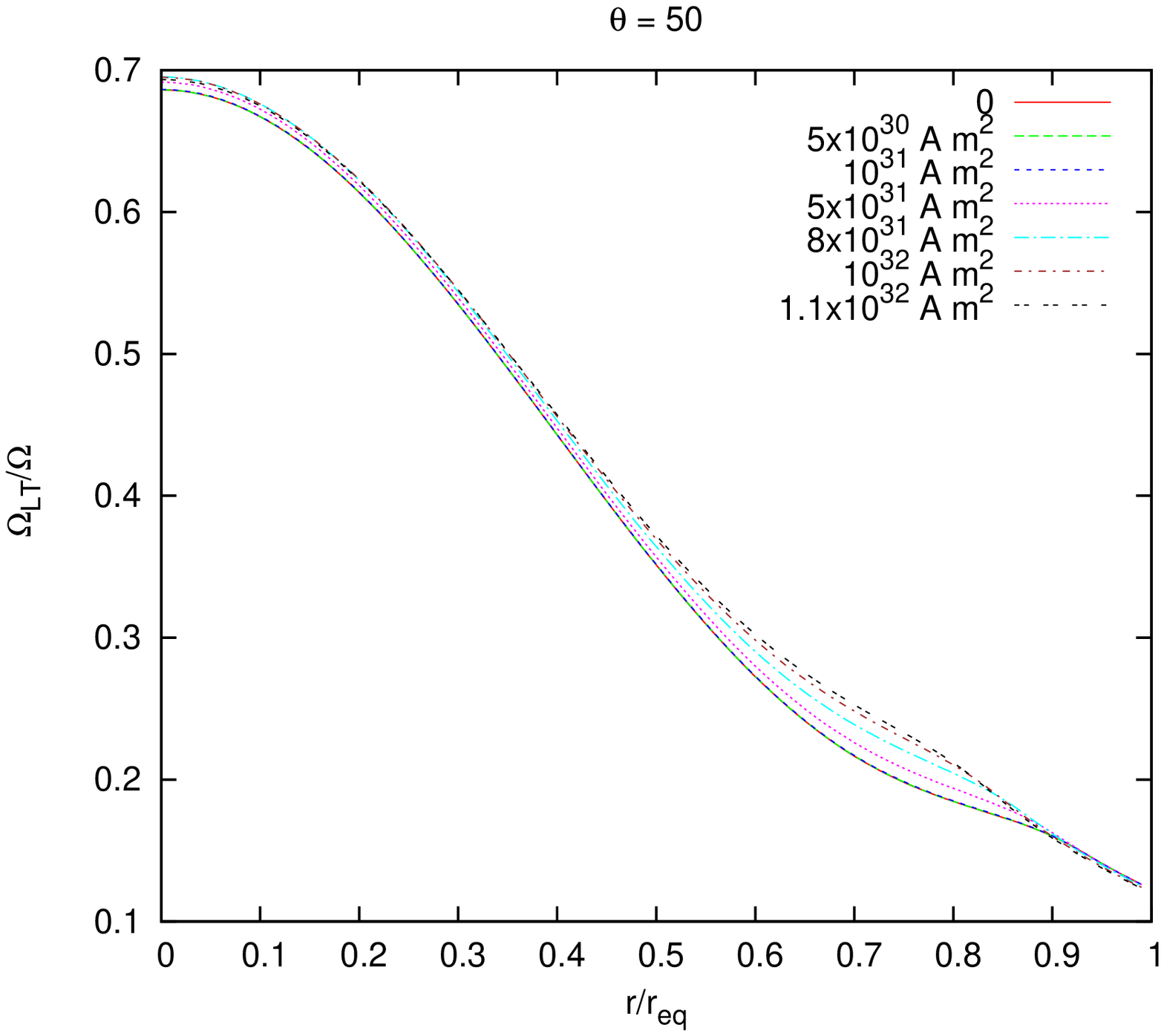}}
 \subfigure[at $70^0$]{
 \includegraphics[width=2.3in,angle=270]{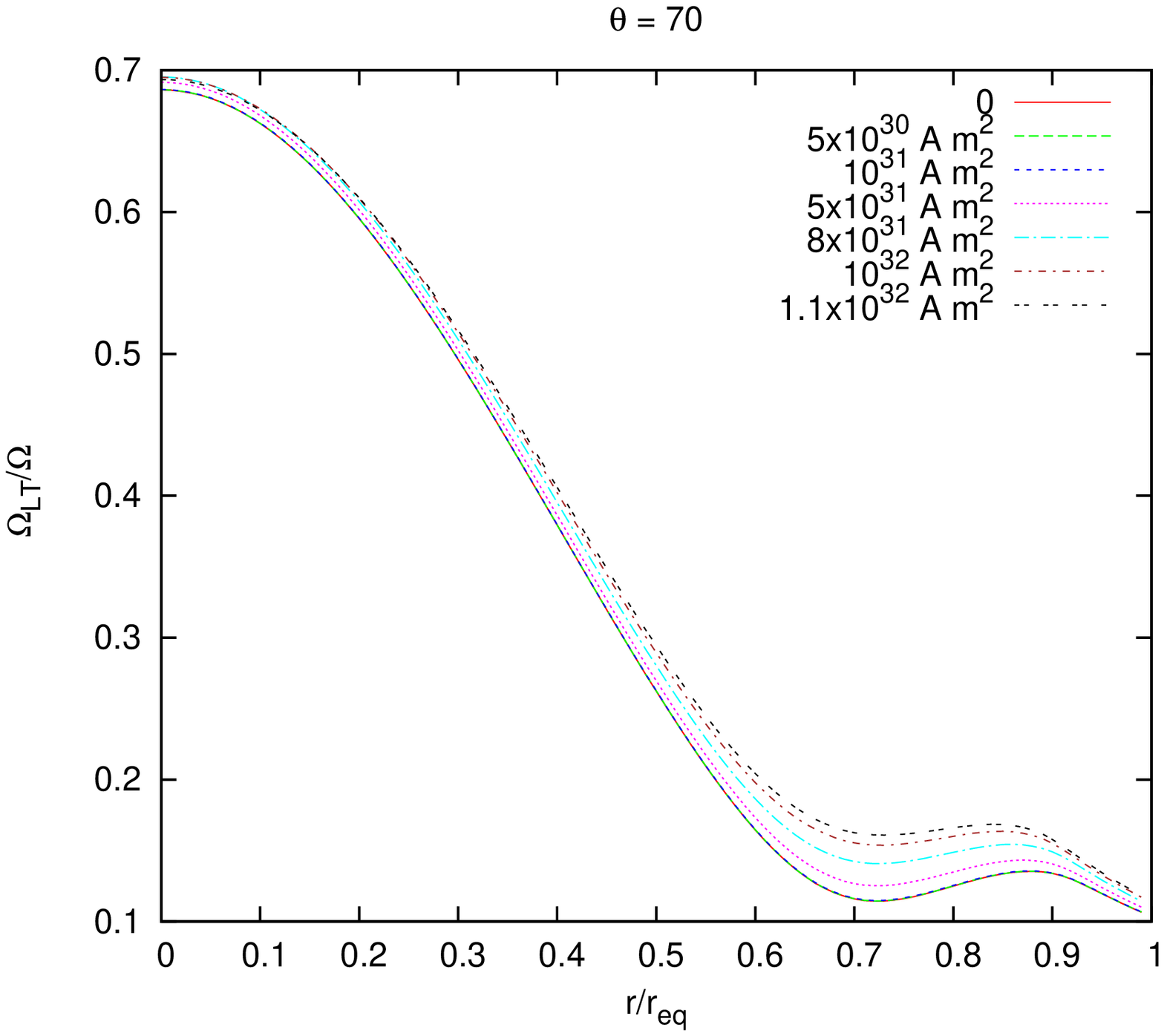}}
  \subfigure[at $90^0$ (along the equator)]{
 \includegraphics[width=2.3in, angle=270]{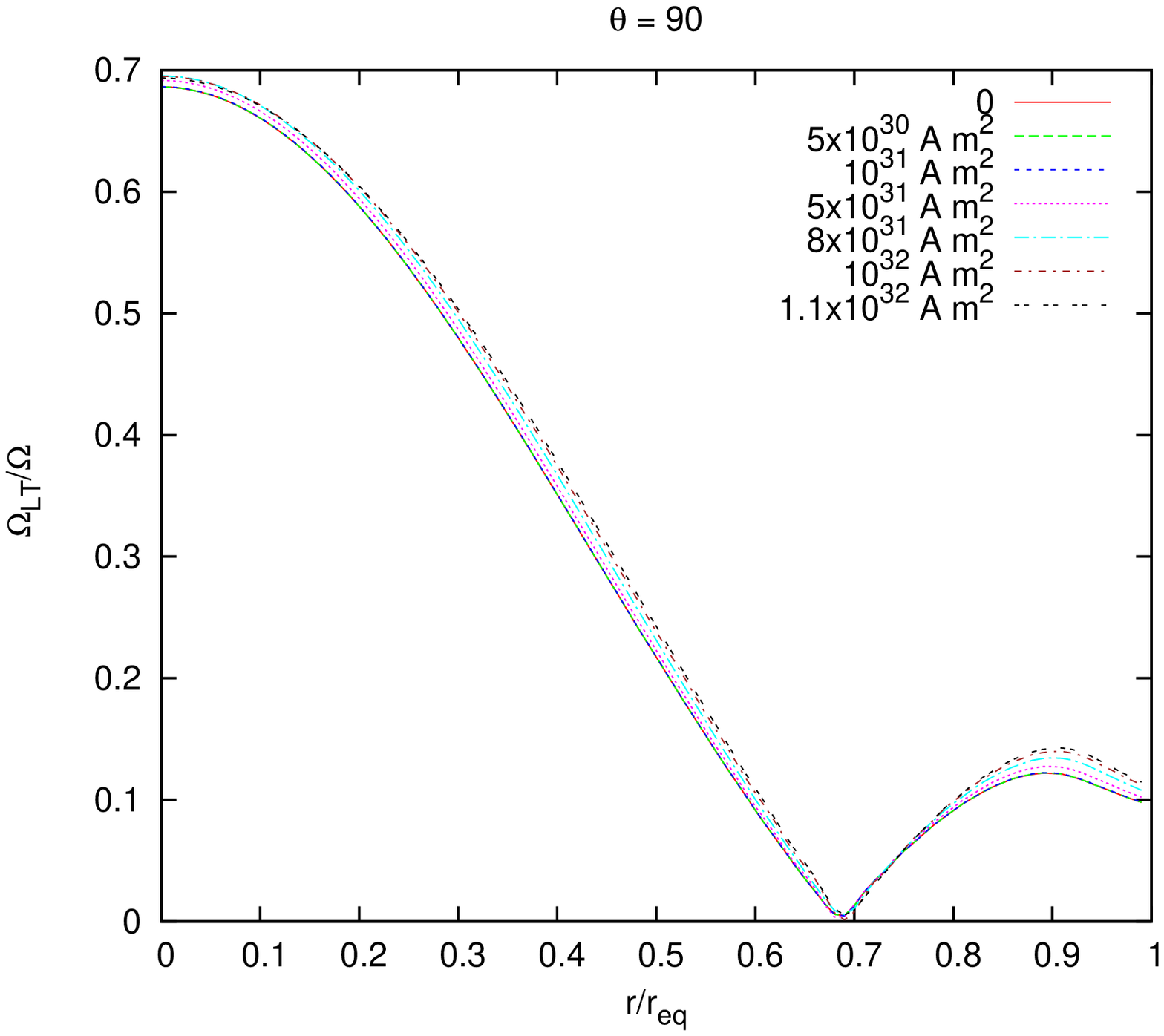}}
\caption{\label{figlt}Plot of normalized frame-dragging rate $\Omega_{LT}/\Omega$ 
inside the pulsar as a function
of radial distance $r$ in units of $r_{eq}$, from the pole ($\theta = 0^0$) to 
equator ($\theta = 90^0$ ), for different values of magnetic moment.}
\label{diffmag}
\end{center}
\end{figure}

Panels (a) to (f) of Fig. \ref{diffmag} reveal that for different 
values of magnetic moment, depending on the angular direction, the frame dragging rate differs inside the pulsar.
Interestingly, the frame dragging
rates along the pole ($\theta = 0^0$) decrease with increasing the magnetic moments and 
the reverse effect is seen along the equator ($\theta = 90^0$ ), i.e., an increase 
in the frame dragging rates with increasing magnetic moments.
Therefore, there should exist a `crossover angle' where the effect
of the magnetic field on the frame-dragging will be `null'.
Magnetic field of the pulsar does not affect the gravitomagnetic effect
in this region, which we can call the `null region'.
To find the null region we show the evolution of the frame-dragging
for various magnetic moments (starting from $0-1.1\times 10^{32}$ A.m$^2$)
from Panel (a) to Panel (f), i.e., from the pole to the equator for every $10^0$
interval. The solid red line represents the zero magnetic moment and
black dashed line represents the highest magnetic moment considered. In Panel (a)
and Panel (f), the deviation between these two lines is maximum which means that 
frame-dragging rate is affected maximally by the magnetic moment of
the pulsar in these two regions. The evolution 
from Panel (a) to (c) shows that the distance between these two lines 
decreases if we proceed from the pole $(0^0)$ to an angle $\theta\sim 40^0$.
Deviation is minimum ($\rightarrow 0$) around $\theta \sim 40^0$ (Panel (c)). This shows that 
magnetism has no effect on frame-dragging at this particular angle. 
If we proceed further from Panel (d) to Panel (f), we see that the deviation
between those two lines (solid red and dashed black) again increases
but in this case it is in the reverse direction.
Thus the maximum effect of magnetic field is along the polar direction 
(reducing the frame-dragging frequency)
and in the equatorial direction (increasing the frame-dragging frequency), 
while for intermediate angles its effect goes through a minimum,
having practically no influence on the frame-dragging frequency around $40^0$.

\subsection{Exterior region ($r \geq r_s$)}
\begin{figure}[tbp]
\begin{center}
\subfigure[at $0^0$ (along the pole)]{
\includegraphics[width=2.3in, angle=270]{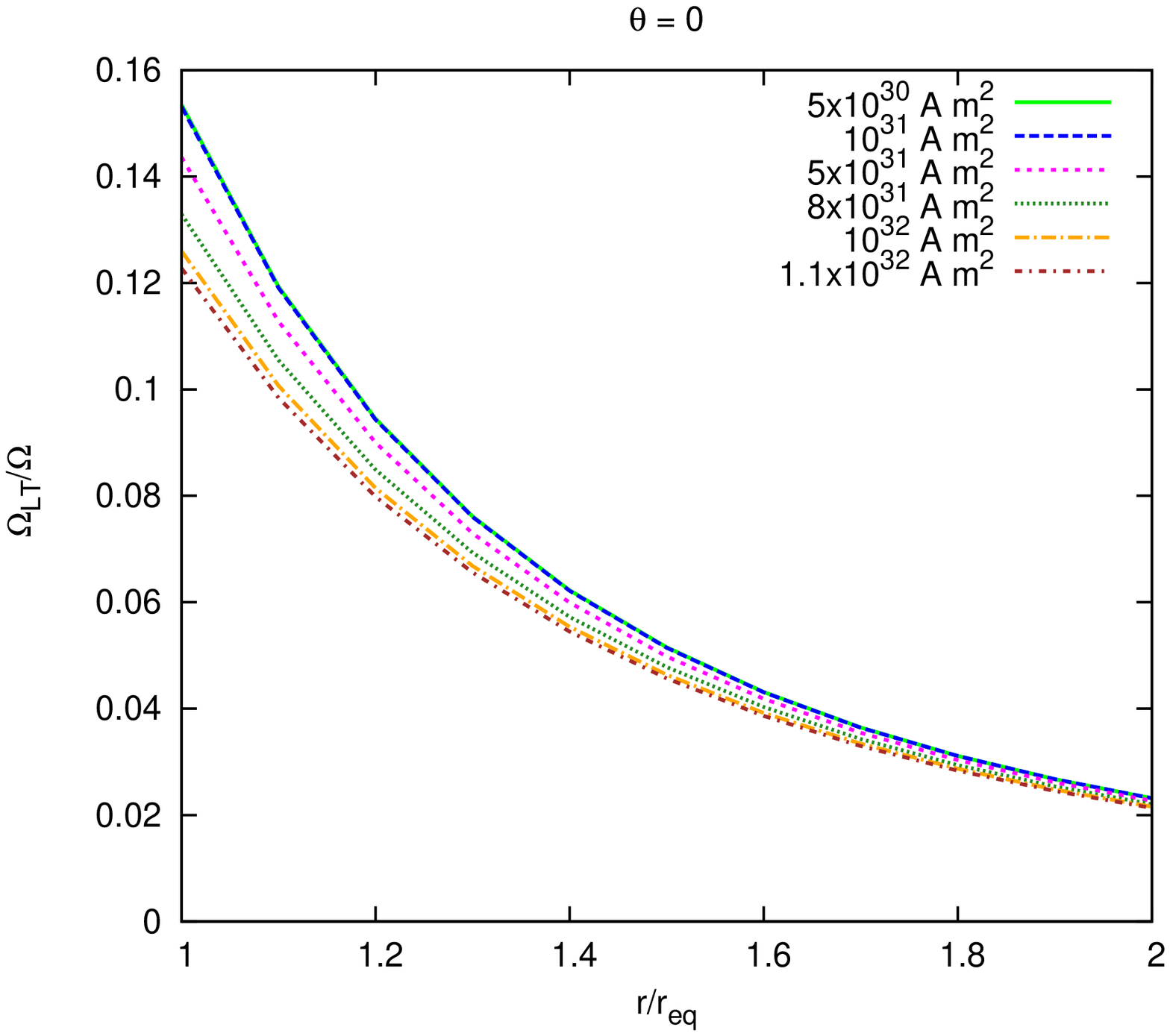}}
 \subfigure[at $50^0$ (along the null Region)]{
\includegraphics[width=2.3in, angle=270]{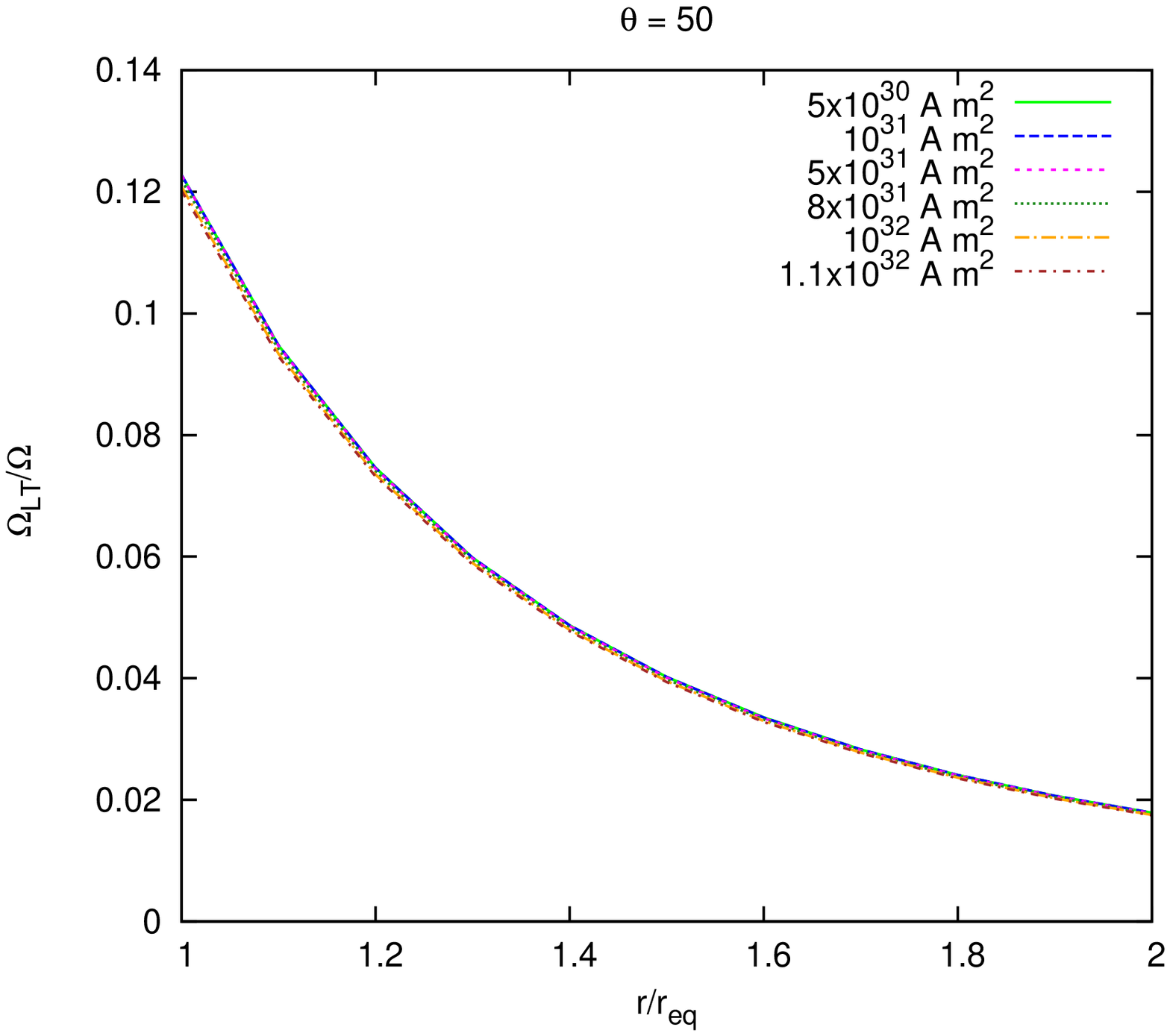}}
 \subfigure[at $90^0$ (along the equator)]{
\includegraphics[width=2.3in, angle=270]{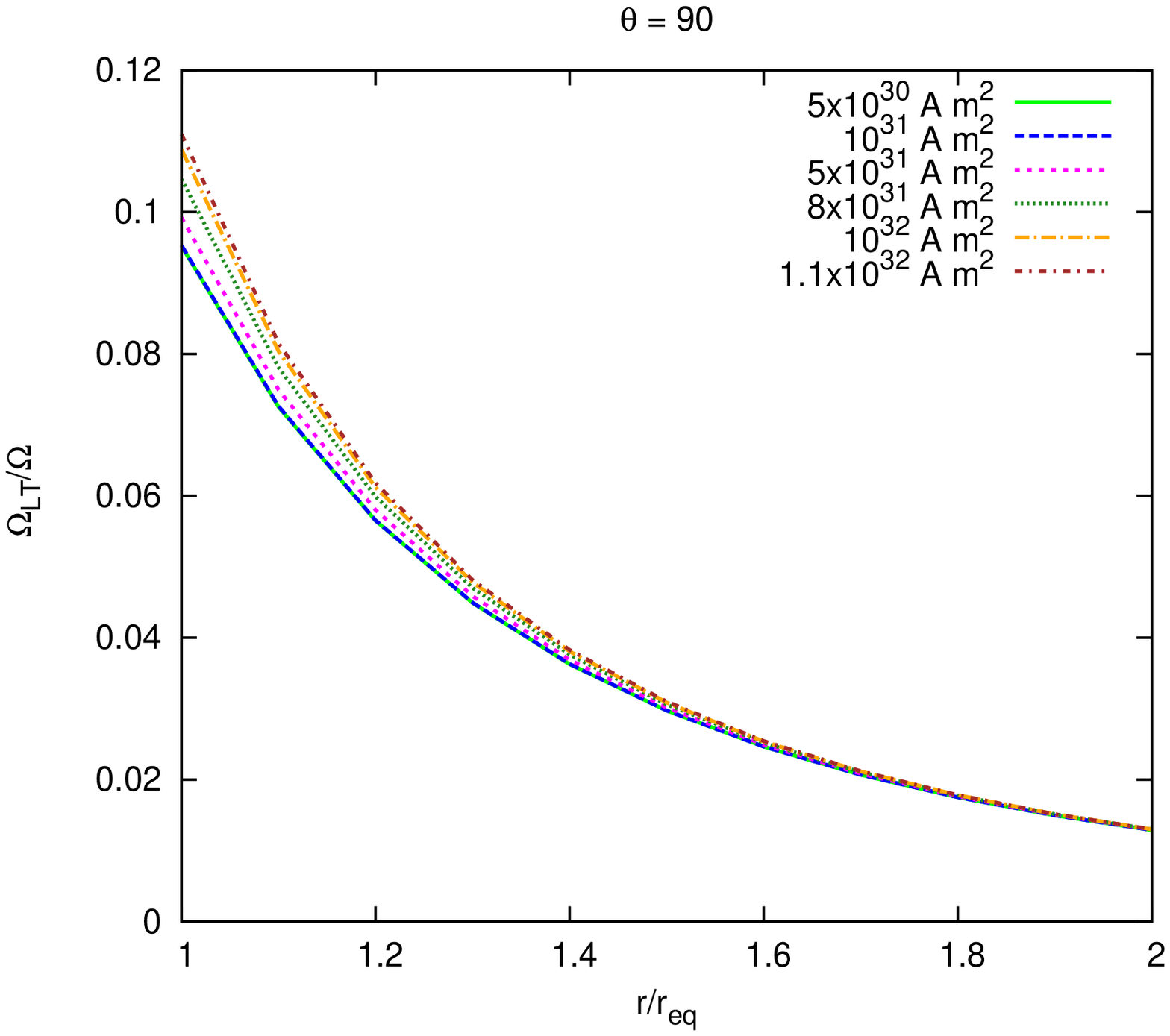}}
\caption{Normalized frame dragging rate $\Omega_{LT}/\Omega$ vs radial distance $r/r_{eq}$ outside the 
neutron star along (a) polar direction, (b) null Region $(50^0)$ 
and (c) equatorial direction for different values of magnetic moment.}
    \label{out}
    \end{center}
  \end{figure}

The frame dragging rates are plotted for radial distances 
extending outside the pulsar ($r \geq r_s$) along polar (panel a), $\theta=50^0$ (panel b) and 
equatorial directions (panel c) in Fig. (\ref{out}), respectively.
This shows that the null region also exists outside the pulsar, at angle of about $50^0$.
This is expected as the effect of the magnetic field 
exists not only inside but also outside (in principle $r\rightarrow \infty$) the pulsar and
the frame-dragging should therefore also extend upto $r\rightarrow \infty$.
The change in the null angle inside and outside can be understood from the fact that, as the energy density
vanishes at the surface of the pulsar, the frame-dragging frequency is no longer determined by the matter solution
where $T^{\mu \nu}$ is given by Eq. (\ref{emtensor}) but by the electrovacuum solution 
with $\varepsilon=P=0$ in $T^{\mu \nu}$. The curves for frame-dragging
frequencies converge asymptotically at large distances ($r >> r_s$). However, there is an interplay between the influence
of the magnetic field on both the matter and vacuum solutions, as well as on the radial distance at which there is a change from
the matter to vacuum solutions.

\begin{figure}[tbp]
\begin{center}
\subfigure[$10^{31}$ A.m$^2$]{
\includegraphics[width=2in, angle=270]{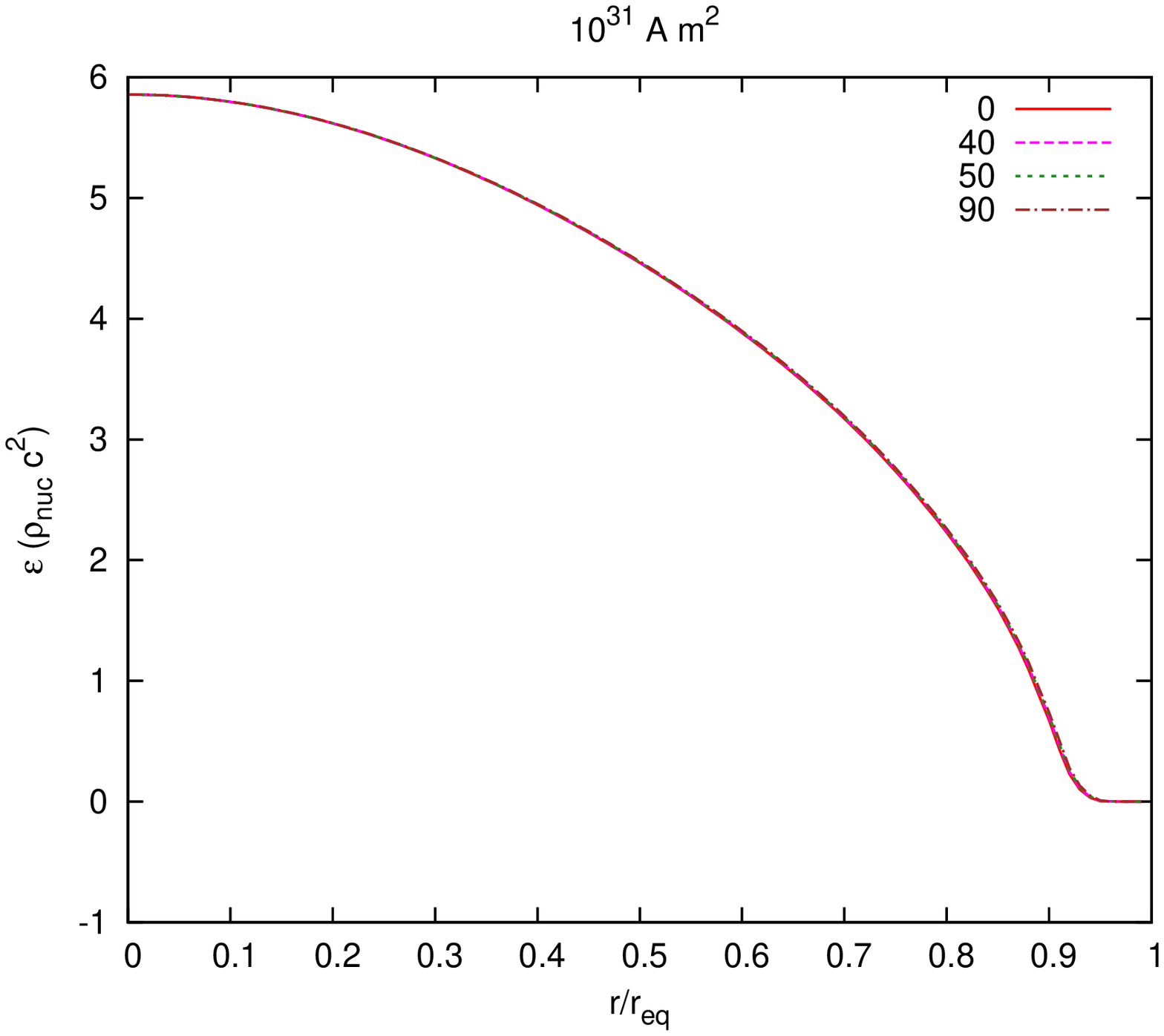}}
\subfigure[$10^{32}$ A.m$^2$]{
\includegraphics[width=2in, angle=270]{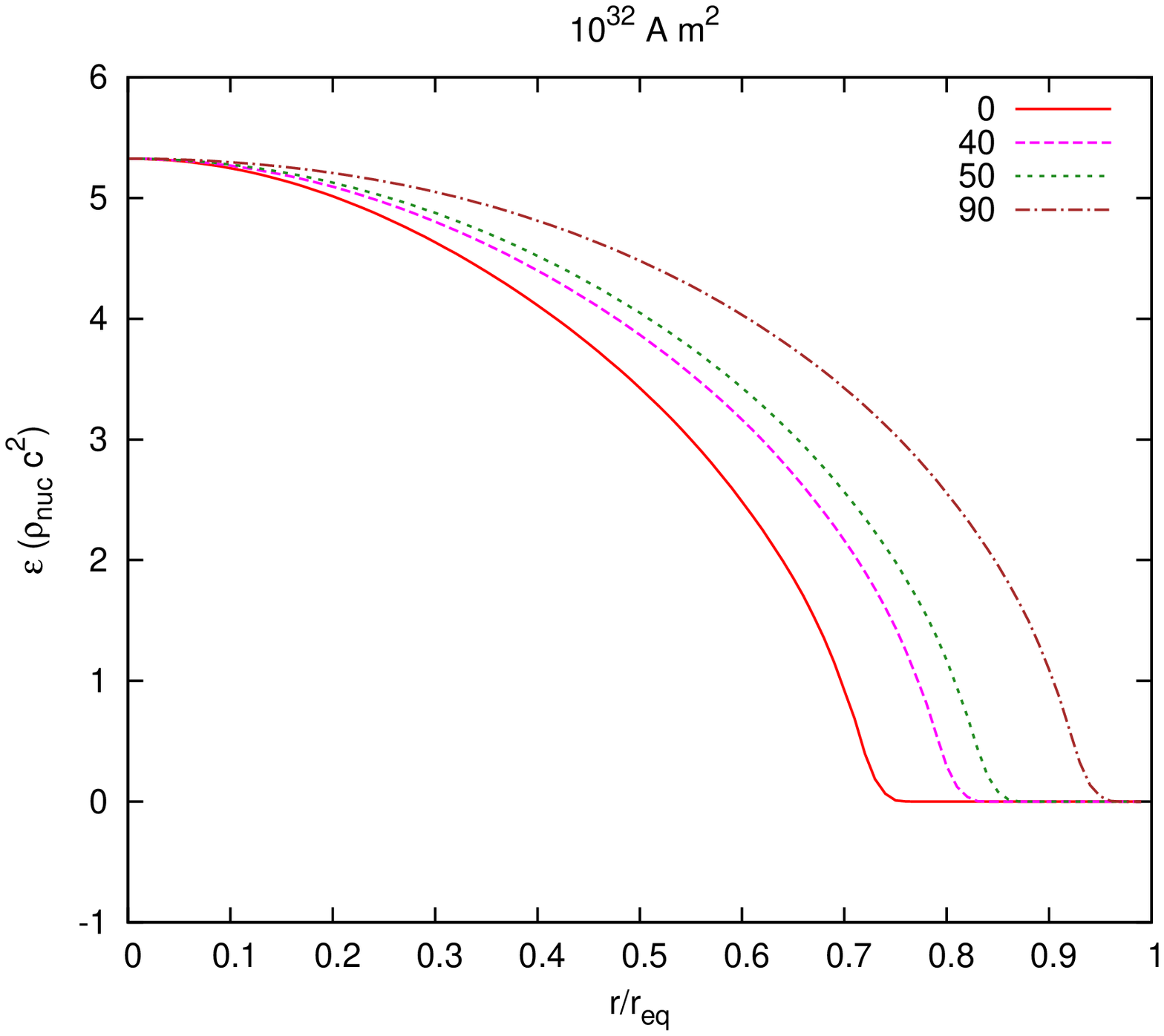}}
\caption{Energy density profile (in units of nuclear density
$\rho_{nuc}$) along polar direction (red solid line), $40^0$ 
(magenta dashed line), at $50^0$ (green dotted line)
and equatorial direction (brown dot-dashed line) for  
magnetic moment of (a) $10^{31}$ A.m$^2$ and (b) $10^{32}$ A.m$^2$.}
    \label{fig:enprof_mm}
    \end{center}
  \end{figure}

Comparing the energy density profiles along different angles for different magnetic moments, we find that for small values
of magnetic moments (e.g. $10^{31}$ A.m$^2$ ) the energy density profiles remain the same along all angular directions
(see Fig. \ref{fig:enprof_mm} (a)). This is expected as 
for small magnetic fields the stellar surface remains spherically
symmetric. However for large magnetic
moments (e.g. $10^{32}$ A.m$^2$ ), the energy density profiles vary 
along different angles (see Fig. \ref{fig:enprof_mm} (b)). 
This can be understood from the fact that
the stellar surface is distorted from spherical symmetry due to the
strong magnetic fields. For a poloidal configuration, 
the deformation of the stellar surface is oblate (polar flattening and equatorial bulging). This means the
energy density goes to zero at smaller radial distances
along the polar direction and at larger radial distances along the 
equatorial direction, defining the distorted stellar surface. It is found that the energy density profile
(and hence the stellar surface) is least affected along $50^0$ 
(comparing Panel (a) and Panel (b) of Fig. \ref{fig:enprof_mm}).

\subsection{Role of the rotation frequency of the pulsar}
\label{sec:rot}
To determine the influence of the value of pulsar rotation on the position of the null region,
we repeat the numerical calculations for $\Omega$ = 0, 100, 1000 $s^{-1}$. As the rotation frequency increases to large values.
the surface of the star is also deformed to a non-spherical shape. However, we found that when
the rotation frequency is zero, the LT frequency vanishes, no matter what the value
of the magnetic field. This implies that there is no contribution of free precession due to the oblateness
caused by the magnetic field, and that the LT effect arises purely 
due to the rotation of the spacetime.
The LT frequency is non-zero only in presence of rotation, and its magnitude is affected
by the presence of the magnetic field. However, we have also observed that the value of the rotation frequency of the pulsar has 
no influence on the value of the angle at which the magnetic field ceases to affect the frame dragging rate. 
For the interior region, the null region is found at $\sim 40^0$
while for the exterior region it is found at $\sim 50^0$, as obtained for $\Omega$ = 276.8 $s^{-1}$
for the PSR J0737-3039.

Here, we should emphasize that any kind of magnetic field is not directly responsible for the 
frame-dragging. This effect solely depends on the rotation of the spacetime
(there is also an exception where LT effect is not related
to the rotating spacetime but some ``rotational sense'' was involved (\cite{cm}) there).
Thus, if a magnetic field exists in a non-rotating spacetime, the spacetime will not
be able to show the frame-dragging. If the spacetime starts to rotate,
gravitomagnetism is induced in this spacetime and the magnetic field
comes into play to show its effect on frame-dragging. 
In an analogy with electromagnetism, we can say 
that as a non-magnetic material is not affected by the magnetic 
field of any magnet, the non-rotating spacetime is also not 
affected by the magnetic field of that spacetime in the case of frame-dragging.
It is evident form Eq.(\ref{eq:metric}) that $N^{\phi}$ vanishes for a 
non-rotating spacetime ($\Omega=0$) and therefore the gravitomagnetic effect 
vanishes ($\Omega_{LT}=0$) which is also evident from Eq.(\ref{eq:omega_lt}),
though the magnetic field is non-zero (see Eq.(\ref{emtensor})).
The {\it tussle between magnetic field and gravitomagnetic effect
is only possible if they are both present in a particular spacetime}.
If any one of them is absent, the tussle will also be absent.
That is why it is meaningless to isolate the role of
magnetic moment on LT effect from that due to rotation only,
but the deviation of the LT effect due to the presence of the magnetic 
field can easily be deduced from the Panels (a)-(f)
of Fig. \ref{diffmag} for various magnetic moment values at various angles.

\subsection{Role of the magnetic field geometry}
\label{sec:geometry}
It is instructive to discuss the role played by the simplifying assumptions that went into the construction of the 
 model for rotating magnetized neutron stars. It was outlined in Sec. \ref{sec:global} that the poloidal field geometry
 considered in this work was intuitive, but definitely not the most general case. Unlike the external field geometry,
 the internal magnetic field configuration cannot be determined by direct observations. It has been suggested that 
 differential rotation in newborn neutron stars could create a strong toroidal magnetic component.
 The effect of purely toroidal fields on the structure of neutron stars has been elaborately studied 
 in ~\cite{Kiuchi, Yasutake, Frieben}. Numerical simulations of purely poloidal and toroidal geometries have found
 them to be unstable \cite{Lander2011,Braithewaite06}. However some studies found rapid uniform rotation 
to suppress instability in both purely poloidal and toroidal configurations.
Among mixed-field configurations, the twisted-torus geometry was found to be the most promising candidate for
a stable magnetic field configuration \cite{Yoshida06, Glampedakis2012, Ciolfi09,Pili2014}. All these models
consider a poloidal-dominated geometry, with toroidal-to-poloidal energy ratio restricted to less than 10$\%$. 
Recently, Ciolfi and Rezzolla \cite{CiolfiRezzolla} succeeded in producing toroidal dominated twisted-torus configurations, but their stability
in nonlinear simulations is still to be established.  

 Although the frame dragging rate depends only on the 
 spacetime metric and not directly on the magnetic field geometry (see Sec. \ref{sec:ltprec}), the interplay of the 
 frame dragging rate with the magnetic field profile should depend on the chosen magnetic field geometry. 
 It has been found that while purely poloidal magnetic fields 
 deform a neutron star shape and matter distribution to an oblate shape (equatorial radius larger than polar radius) as 
 in Fig. \ref{fig:xzcut}, purely toroidal fields fields deform them to a prolate shape. On the other hand, rotation induces oblateness in the 
 surface deformation. Thus for rotating magnetized neutron stars with a purely toroidal geometry, different combinations
 of the surface deformation are possible depending on the relative strength of the field and rotation rate \cite{Frieben}.
 Thus to summarize, the matter distribution of the star can vary with the magnetic field geometry in a very complicated way,
 and this would affect its interplay with frame-dragging. But this is beyond the scope of this work and we leave the general case
 for future study. However it must be outlined here that for both purely poloidal and toroidal cases,
 the stellar shape would be least distorted close to an angle $40-50^0$, and there should exist a null region following the same
 arguments. 
 
\section{ Measurement of frame dragging in neutron stars }
\label{secnull}
In Sec. \ref{sec:rot}, we discussed the 
LT effect of PSR J0737-3039 and without making any 
realistic experiment we predicted that the null region 
will be at $\sim 40^0$ (inside the pulsar) and $\sim 50^0$ (outside the pulsar) where 
magnetic field has no effect on 
gravitomagnetism, whatever magnetic field value PSR J0737-3039 possesses. 
The crossover angle of the null Region between $40^0$ to $50^0$
occurs on the surface of the pulsar at $\sim 0.8r_{eq}$  
which is evident from Panels (c)-(d) of Fig. \ref{diffmag} and Panel (b) of Fig. \ref{out}.
It is also evident from Panel (b) of Fig. \ref{fig:enprof_mm} that 
the distance of the surface of the pulsar is $r_s \sim 0.8~r_{eq}$  at $40^0$ angle.

\begin{figure}[tbp]
    \begin{center}
\includegraphics[width=3.0in]{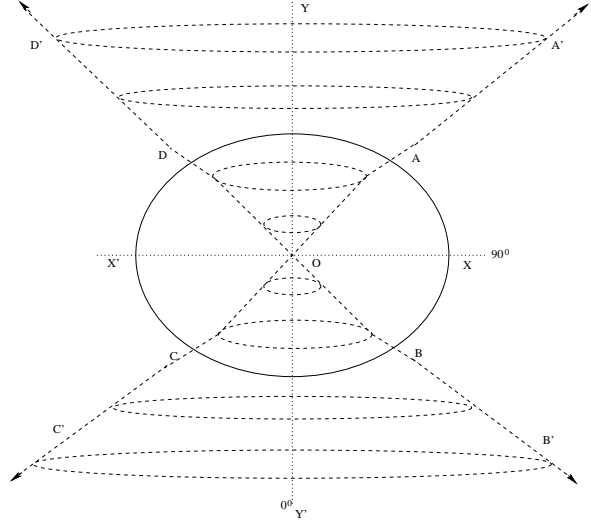}
      \caption{\label{fignl} Azimuthal section of a {\it theoretically predicted} null Region/Surface
      of the pulsar PSR J0737-3039 around the symmetry axis $YOY'(0^0)$ or
      the rotation axis of the pulsar. $X'OX$ is the equatorial plane and solid
      black line represents the surface of the pulsar. $2\theta_{in}=\angle DOA=
     \angle BOC\approx 2\times40^0=80^0$. The boundary of the 
3-D hypothetical hollow `cone' has been drawn 
along $D'DOAA'$ and $B'BOCC'$ around the symmetry axis. 
In principle, the cone is extended upto $r\rightarrow \infty$, which is shown
by four arrows but outside 
i.e., $ 0.8r_{eq} < r \leq \infty$ (as $OA=OB=OC=OD \approx0.8r_{eq}$),
the null region appears at $\sim 50^0$ instead of $40^0$ which is evident from
Panels (c)-(d) of Fig. \ref{diffmag} and Panel (b) of Fig. \ref{out}.
If the gyro moves along the surface of this 3-D hypothetical `cone', 
the effect of  {\it magnetic field} of the pulsar will not be exerted on its 
frame-dragging frequency or gravitomagnetic effect.
Inside this cone, $\Omega_{LT}$ decreases with increasing magnetic moment
and it shows completely reverse effect outside this cone
(this figure is not drawn in scale).}
      \end{center}
\end{figure}

In an axisymmetric spacetime, we can imagine that the null angle
$2\theta_{in} \sim 2\times40^0=80^0$ 
creates a hypothetical 3-D hollow `cone' inside the pulsar and
$2\theta_{out} \sim 2\times50^0=100^0$ outside, which is depicted in Fig. \ref{fignl}.
Therefore, if a test gyro moves 
along the surface of the cone, LT frequency of the gyro will not be affected
by the magnetic field of the pulsar whatever magnetic field it holds.
If the gyro moves along any direction inside the cone, its LT frequency 
varies inversely with the magnetic moment and outside the cone the 
frequency will be directly proportional to the magnetic moment. 

Now, using the concept of Gravity Probe B 
experiment we can suggest a thought experiment 
which can be performed by a test gyro. We can make it move towards 
the strong gravitational field of a pulsar along the pole or equator
or whichever angle we want. For a particular angle outside an unknown pulsar, say, $\theta=\theta_d$, 
we make the gyro move from infinity towards the pulsar and in principle we could measure the 
exact frame-dragging rate. We could then calculate the difference between
this measured frame dragging rate 
and the theoretically calculated frame dragging rate for zero magnetic field. 
The difference in the frame dragging along this given angular direction will be determined by the magnetic field of the pulsar. 
Now, we can move the gyro along different angles $\theta_d$ from $0^0$ to $90^0$ and 
the angle at which the measured frame dragging frequency is exactly 
equal to the calculated LT precession
frequency for zero magnetic field, is the null angle ($\theta_d=\theta_{out}$) outside the 
pulsar. As an example, it is $50^0$ in Fig. \ref{fignl}. 
In this way, we can determine the null angle 
for the effect of magnetic field on LT precession for the particular pulsar.
After determining the $\theta_{out}$ we can move the gyro further toward the 
unknown pulsar. Now, very close to the pulsar surface (point A, B, C or D in Fig. \ref{fignl}) we will see that the
frame-dragging rate is starting to get affected by the magnetic field suddenly.  
Now, we have to follow the similar procedure (which we have followed outside the pulsar)
at the surface of the neutron star to find the null angle 
$\theta_{in}$ ($40^0$ in Fig. \ref{fignl}) inside the pulsar.

With the rapid advancement in observational facilities, other possibilities are emerging
that may allow the measurement of LT precession in neutron stars in the near future.
It has been reported recently that with sufficient increased timing measurements and new generation telescopes 
such as Square Kilometre Array (SKA), LT precession of the orbit of PSR J0737-3039 will become measurable \cite{Kehl}.
This also opens up the possibility of constraining neutron star
EoSs (see \cite{Kehl} for a detailed discussion).

Another interesting possibility would be to use pulsar planets as gyroscopes
around neutron stars to study frame dragging effect. About five or more 
planets have been discovered orbiting pulsars (e.g. PSR B1257+12, PSR B1620-26, 
PSR J1719-1438) \cite{Bailes,Wolszczan}.
LT precession rate of the planet for PSR J1719-1438 can be calculated
easily using the weak-field formulation (as $r >> GM/c^2$) of LT effect :
$\O_{\rm LT} \sim GMR^2\O/(c^2r^3)$ where mass of the pulsar
$M = 1.4M_{\odot}$, radius $R \approx 10$ km, angular velocity
$\O=2\pi/(5.7 \times 10^{-3})$ rad.s$^{-1}$, semi-major axis 
$r \sim 0.004$ AU \cite{Bailes2}, gravitational constant $G=6.67\times 10^{-11}$ N.m$^2$.kg$^{-2}$
and speed of the light in vacuum $c=3\times 10^8$ m.s$^{-1}$. It is found to be
approximately $10^{-12}$ rad s$^{-1}$, i.e., after one year 
($\sim 10^7$ s) the spin of the planet precesses 
approximately by $10^{-5}$ radian ($\approx 10^3$ mas/yr
or $ 10^{-4}$ degree/yr). This is a small effect, but much larger in comparison
with the case of Gravity Probe B experiment, where
LT frequency is measured to be about 39 mas/yr due to the rotation of the earth.
Thus, although it would be difficult to measure the LT effect
in a pulsar-planet system, it should in principle be measurable over 
a sufficient long period of observation ($\sim$ 20 years).
This observation of 20 years is meant for the present day radio
telescopes but with the highly sensitive radio telescope such as SKA, it
would be much less.

\section{Summary and Outlook}
\label{seccon}

In this work, we extended the recent calculation \cite{chandra} for exact
LT precession rates inside and outside neutron stars to include
effects of electromagnetic field. This should be particularly
interesting for the case of high magnetic field pulsars and magnetars.

We employed a well established numerical scheme \cite{BGSM,chatterjee,Bocquet}
to obtain equilibrium models of fully relativistic magnetized
neutron stars with poloidal geometry. We computed the frame dragging rates inside and outside 
the stellar surface for such models. We obtained qualitatively similar
peculiar features for the frame dragging rates along polar and
equatorial directions as previously discussed for non-magnetic calculations \cite{chandra}, with the effect
of the electromagnetic field evident on the frame dragging
rate inside as well as outside the star. This could be potentially interesting for high magnetic field pulsars to
constrain the magnetic field from a measurement of the LT frequency. As an example for PSR J0737-3039, we found that 
the maximum effect of magnetic field appears along the polar direction (decreasing the frame-dragging frequency)
and in the equatorial direction (increasing the frame-dragging frequency), while for intermediate angles 
its effect goes through a minimum, having practically no influence on the frame-dragging frequency around $40^0$ inside
and $50^0$ outside the pulsar. We showed that this can be attributed to the twofold effect of the magnetic field, on the LT frequency
as well as on the breaking of spherical symmetry of the neutron star. We also devised a thought-experiment to determine
the null angle using a measurement of its LT frequency.

Measurements of geodetic and frame dragging effects by dedicated 
missions such as LAGEOS and Gravity Probe B \cite{Overduin} have permitted us to 
impose constraints and better understand gravitomagnetic-magnetic aspects
of theories of gravity. There is a need to expand the sensitivity of future 
gravitational tests to neutron stars to understand phenomena such as LT precession.
We have discussed briefly the possibility of measuring frame dragging
in neutron stars in the future.

\section*{Acknowledgments}
DC is grateful to J\'erome Novak for his guidance regarding the numerical computation. 
CC gladly acknowledges Kamakshya P. Modak and Prasanta Char for their
valuable suggestions. The authors sincerely thank the anonymous
referee for the valuable comments and suggestions that helped improve the manuscript.

\end{document}